\documentclass[]{spie}  

\usepackage[]{graphicx}
\usepackage{longtable}

\usepackage{amsmath,amsfonts,amssymb}
\usepackage{graphicx}
\usepackage[colorlinks=true, allcolors=blue]{hyperref}

\title{Observing the PTPS sample of evolved exoplanet host candidates using the NPOI}

\author[a]{Ellyn K. Baines}
\author[a]{J. Thomas Armstrong}
\author[a]{Henrique R. Schmitt}
\author[b]{R. T. Zavala}
\author[b]{James A. Benson}
\author[c]{Andrzej Niedzielski}
\author[d]{Pawel Zielinski}
\author[e]{Martin Vanko}
\author[f]{Aleksander Wolszczan}

\affil[a]{Naval Research Laboratory, Washington, DC, USA}
\affil[b]{U.S. Naval Observatory, Flagstaff Station, Flagstaff, AZ, USA}
\affil[c]{Nicolaus Copernicus University, Torun, Poland}
\affil[d]{Department of Theoretical Physics \& Astrophysics, Masaryk University, Brno, Czech Republic}
\affil[e]{Astronomica Institute, Slovak Academy of Sciences, Tatranska Lomnica, Slovakia}
\affil[f]{Pennsylvania State University, University Park, PA, USA}

\authorinfo{Further author information: (Send correspondence to E.K.B.)\\E.K.B.: E-mail: ellyn.baines@nrl.navy.mil, Telephone: 1 202 767 2437}

\begin{document} 
\maketitle

\begin{abstract}
We plan to measure the angular diameters of a sample of Penn State-Torun Planet Search (PTPS) giant exoplanet host star candidates using the Navy Precision Optical Interferometer. The radii of evolved giant stars obtained using spectroscopy are usually ill-defined because of the method's indirect nature and evolutionary model dependency. The star's radius is a critical parameter used to calculate luminosity and mass, which are often not well known for giant stars. Therefore, this problem also affects the orbital period, mass, and surface temperature of the planet. Our interferometric observations will significantly decrease the errors for these parameters. We present preliminary results from NPOI observations of six stars in the PTPS sample.
\end{abstract}



\section{INTRODUCTION}

\hskip15pt One of the challenging topics of modern astronomy is the search for exoplanets. Many exoplanets known today are gas giants, close to their central star and detected using the radial velocity (RV) and photometric transit techniques. The orbital properties of these planets and their minimum masses are calculated using the mass ($M$) of the central star. $M$ relies on evolutionary tracks and is calculated from the luminosity ($L$) of the star, which is proportional to the radius ($R$) squared. Because $R$ is usually not well known for giant stars, this parameter should be measured directly when possible. This can be accomplished using interferometric observations, which measure the star's angular diameter.\cite{2001ARAandA..39..353Q} That value, when combined with distance measurements from parallax studies, then produces a physical radius for the star.

Our observing list is a subset of the Penn State-Torun Planet Search (PTPS) survey\cite{2008IAUS..249...43N} of giant stars with RV planet candidates. The survey consists of 744 evolved stars, 455 of which have a complete spectroscopic analysis. Niedsielski et al.\cite{2016AandA...589C...1N} used measured spectroscopic properties, such as surface gravity, effective temperature, and metallicity, combined with a distance measurement to derive $R$ and $L$. Then they estimated $M$ and age using the evolutionary tracks and isochrones. Unfortunately this method is uncertain due to difficulties in accurate placement of a single star on red giant clump, which is a very complex region on H-R diagram. We plan to improve this by direct measurements of stellar sizes. 

Moreover, together with a high-precision interferometric radii determinations and rotational velocities obtained from high resolution spectra, we can analyze stellar surface inhomogeneities such as spots using RV variations. The precise rotation period is needed to rule out stellar atmospheric activity as the source of such variations. Hence, a detailed knowledge of stellar fundamental parameters is very important for interpretation of the giant stars' characteristics in the context of exoplanet searches.

Based on the interferometric measurements, we will obtain better estimates of $M$. If we can determine the angular diameter with a precision of 2$\%$, $R$ can be determined with a precision of better than 5$\%$. The largest error source is in the parallax,\footnote{This is the case until the GAIA mission\cite{2001AandA...369..339P} starts releasing improved parallax measurements. Then the errors on stellar radii will be correspondingly more precise. Having an existing archive of interferometrically measured angular diameters will be useful to quickly produce radii using the GAIA parallaxes when they become available.} so, after taking into account a 10$\%$ precision, the error on $M$ will then be approximately 0.1 $M_{\odot}$. This is significantly better when compared to previous estimations based on moderate quality photometry and evolutionary models. For the PTPS targets, we predicted mass values from 1 to 3 $M_{\odot}$ and, even after critical assessment, an uncertainty of 30$\%$ remains possible. Hence, the interferometric data will improve the precision of $M$ sufficiently for the subsequent analysis.

The NPOI is capable of observing 25 of the PTPS sample stars after culling the list based on estimated angular diameters (0.8 mas or larger) and brightness ($V$ magnitude of 5.5 or brighter). We have so far obtained data on six of those stars, and present the preliminary findings here. Section 2 describes the interferometer and observing sequence, and Section 3 discusses how we determine angular diameters, radii, bolometric flux, temperatures, and masses from the combination of interferometric data and photometric values from the literature, and Section 4 summarizes our results.


\section{INTERFEROMETRIC OBSERVATIONS}

\hskip15pt The NPOI is an optical interferometer located on Anderson Mesa, AZ, and is a joint project of the Naval Research Laboratory and the United States Naval Observatory in partnership with Lowell Observatory. It has been in operation since 1994, when the first 18-m and 36-m baselines came into operation \cite{1998ApJ...496..550A}. The NPOI consists of two nested arrays: the four stations of the astrometric array (AC, AE, AW, and AN, which stand for astrometric center, east, west, and north, respectively) and the ten stations of the imaging array, of which six stations are currently in operation (E3, E6, E7, W4, W7, and N3) and four more will be coming online in the near future (N6, N7, E10, and W10).\footnote{The NPOI consists of three arms with general north, east, and west orientations, and each arm has ten stations with 1 being closest to the center of the array and 10 being the farthest out. The baseline is the distance between two imaging elements.}

The current baselines range from 9 to 98 m, and our maximum baseline will be 432 m when the E10 and W10 stations are completed. We use a 12-cm region of the 50-cm siderostats and observe in 16 spectral channels spanning 550 to 850 nm. We combine light from up to six elements at a time to perform baseline bootstrapping, a method in which we create a chain of connected short baselines, where the signal is strong and easily detected, to build up to a long baseline, where the signal is weak but where there is maximum information about the details of a star. The long-baseline observations tell us about asymmetries due to rapid rotation, gravity and limb-darkening effects, star spots, and previously undetected companions.

The observed quantity of an interferometer is fringe contrast or ``visibility'' ($V$). Each NPOI observation was made using the Classic beam combiner and consists of a 30-s coherent (on the fringe) scan in which the fringe contrast is measured every 2 ms, paired with an incoherent (off the fringe) scan used to estimate the additive bias affecting the visibility measurements \cite{2003AJ....125.2630H}. The reduction and analysis of the data are divided into three parts: in the first stage, we reduce the raw visibility data from the 2-ms frames to 1-s averages, removing the empirically determined detector bias in the process. The dispersion of 1--second points provide an estimate of the internal uncertainties. 

The second and third stages take place within the data reduction package $Oyster$.\footnote{Written and maintained in IDL by Christian Hummel, http://www.sc.eso.org/$\sim$chummel/oyster/oyster.html} In the second stage, the data are edited for delay-line errors and other anomalies, background flux is subtracted, scan-average visibility data and uncertainties are calculated, and the data are calibrated. In the third stage we fit model parameters to the calibrated data. The sophistication of all three parts has continued to grow over the lifetime of the NPOI. The range and complexity of models within $Oyster$ is quite broad, including in particular multiple systems of stars with differing colors, limb-darkened disks (using the coefficients of van Hamme\cite{1993AJ....106.2096V}), and gravity darkened rapid rotators.

We observed six PTPS stars over a period of time stretching from May 2007 to May 2016. For each target, we interleaved data scans of the stars with one or two calibrator stars. Our calibrators are stars that are significantly less resolved on the baselines used than the target stars, which meant that uncertainties in the calibrator's diameter did not affect the target's diameter calculation as much as if the calibrator star had a substantial angular size on the sky. The calibrator and target scans were measured as close in time and angular separation as possible, which allowed us to convert instrumental target and calibrator visibilities to calibrated visibilities for the target.

To estimate the calibrator stars' angular diameters, we created spectral energy distribution (SED) fits. We used Kurucz model atmospheres\footnote{Available to download at http://kurucz.cfa.harvard.edu.} based on effective temperature ($T_{\rm eff}$) and surface gravity (log~$g$) values from the literature and fit them to observed photometry after converting magnitudes to fluxes using Colina, Bohlin $\&$ Castelli\cite{1996AJ....112..307C} for $UBVRI$ values and Cohen, Wheaton, $\&$ Megeath\cite{2003AJ....126.1090C} for $JHK$ values (see Table 1). This also allowed us to check if there was any excess emission that might be due to an otherwise unknown low-mass companion or circumstellar disk. Any calibrator candidates displaying variable radial velocities, photometric variations, or any indication of binarity were discarded. 


\begin{center}
Table 1. Calibrators \\
\begin{scriptsize}
\vspace{0.01in}
\begin{longtable}{ccccccccccccc} \hline\hline
\label{example}
\vspace{0.01in}
   & Calib & $U$   & $B$   & $V$   & $R$   & $I$   & $J$   & $H$   & $K$   & $T_{\rm eff}$ & log $g$       & $\theta_{\rm est}$ \\
HD & For   & (mag) & (mag) & (mag) & (mag) & (mag) & (mag) & (mag) & (mag) & (K)           & (cm s$^{-2}$) & (mas)              \\
\hline \endfirsthead
\caption[]{\emph{continued}} \\
\hline
   & Calib & $U$   & $B$   & $V$   & $R$   & $I$   & $J$   & $H$   & $K$   & $T_{\rm eff}$ & log $g$       & $\theta_{\rm est}$ \\
HD & For   & (mag) & (mag) & (mag) & (mag) & (mag) & (mag) & (mag) & (mag) & (K)           & (cm s$^{-2}$) & (mas)              \\
\hline
\multicolumn{13}{r}{\emph{continued on the next page}}
\endfoot
\hline \endlastfoot
\hline
87696  &  90537 & 4.74 & 4.67 & 4.49 & 4.38 & 4.29 & 4.27 & 4.05 & 4.00 &  7943 & 4.27 & 0.56$\pm$0.03 \\ 
89021  &  90537 & 3.54 & 3.48 & 3.45 & 3.40 & 3.40 & 3.44 & 3.46 & 3.42 &  8913 & 3.84 & 0.74$\pm$0.04 \\ 
118098 & 113226 & 3.60 & 3.49 & 3.37 & 3.31 & 3.25 & 3.26 & 3.15 & 3.22 &  8511 & 4.19 & 0.81$\pm$0.03 \\ 
120136 & 113226 & 5.03 & 4.98 & 4.50 & 4.21 & 3.96 & 3.62 & 3.55 & 3.51 &  6456 & 4.25 & 0.82$\pm$0.02 \\ 
166014 & 161797 & 3.77 & 3.81 & 3.84 & 3.85 & 3.89 & 3.97 & 3.96 & 3.95 & 10590 & 4.17 & 0.52$\pm$0.02 \\ 
184006 & 181276 & 4.04 & 3.93 & 3.79 & 3.68 & 3.62 & 3.74 & 3.69 & 3.60 &  8180 & 4.29 & 0.70$\pm$0.04 \\ 
195810 & 188512 & 3.44 & 3.91 & 4.03 & 4.07 & 4.18 & 4.66 & 4.55 & 4.38 & 13600 & 2.44 & 0.39$\pm$0.04 \\ 
217891 & 219615 & 5.20 & 4.62 & 3.70 & 3.17 & 2.71 & 2.02 & 1.49 & 1.39 &  5012 & 2.74 & 0.29$\pm$0.01 \\ 
\hline
\end{longtable}
\end{scriptsize}
\end{center}
\vspace{-0.3in}
$UBV$ values from Mermilliod\cite{2006yCat.2168....0M} except for HD 120136 (from Morel \& Magnenat\cite{1978AandAS...34..477M}) and HD 184006 (from Johnson et al.\cite{1966CoLPL...4...99J}); $RI$ values from Johnson et al.\cite{1966CoLPL...4...99J}; $JHK$ from Cutri et al.\cite{2003tmc..book.....C}; $T_{\rm eff}$ and log $g$ from Allende Prieto \& Lambert\cite{1999AandA...352..555A} except for HD 166014 (from Wu et al.\cite{2011AandA...525A..71W}) and HD 184006 and HD 195810 (from Cox\cite{2000asqu.book.....C} based on the stars' SIMBAD spectral types).
$\theta_{\rm est}$ is the estimated diameter calculated as described in Section 2.


\section{RESULTS: DETERMINING STELLAR PROPERTIES}

\hskip15pt An interferometer samples part of the Fourier transform of the brightness distribution on the sky. In the simple case of a circular, uniformly bright stellar disk (UD) at the center of the interferometer's field of view, this Fourier transform is $V = [2 J_1(x)] / x$, where $J_1$ is a Bessel function of the first kind and first order and $x = \pi \nu \theta_{\rm UD}$, where $\theta_{\rm UD}$ is the apparent UD angular diameter of the star and $\nu$ is the spatial frequency, measured in units of radians$^{-1}$ \cite{1992ARAandA..30..457S}. Frequency is given by $\nu$ = $B$/$\lambda$, where $B$ is the projected baseline at the star's position and $\lambda$ is the effective wavelength of the observation.

Angular diameter measurements are interpreted as spatial frequencies sampled by the interferometer. The square of the Fourier transform, $[2 J_1(x)/x]^2$, which is typically the quantity measured ($V^2$), is easily visualized: this is the familiar Airy pattern produced by an unresolved star at the focal plane of a circular lens. They are equivalent because a point source produces a uniform $E$ field across the circular pupil, and the lens produces the Fourier transform of the focal input.

A more realistic model of a star's disk includes limb darkening (LD).  If a linear LD coefficient $\mu_\lambda$ is used,
\begin{equation}
V^2 = \left( {1-\mu_\lambda \over 2} + {\mu_\lambda \over 3} \right)^{-1}
\times
\left[(1-\mu_\lambda) {J_1(x_{\rm LD}) \over x_{\rm LD}} + \mu_\lambda {\left( \frac{\pi}{2} \right)^{1/2} \frac{J_{3/2}(x_{\rm LD})}{x_{\rm LD}^{3/2}}} \right] .
\end{equation}
where $x_{\rm LD} = \pi B\theta_{\rm LD}\lambda^{-1}$ (from Hanbury Brown et al.\cite{1974MNRAS.167..475H}). We used $T_{\rm eff}$ and log $g$ values from the literature (from Allende Prieto \& Lambert\cite{1999AandA...352..555A} for all but HD 161797, which was from Valenti \& Fischer\cite{2005ApJS..159..141V}) with a microturbulent velocity of 2 km s$^{\rm -1}$ to obtain $\mu_\lambda$ from Claret \& Bloeman\cite{2011AandA...529A..75C}. The resulting $\theta_{\rm LD}$ and other observed stellar properties are listed in Tables 2 and 3. Figure 1 shows the $\theta_{\rm LD}$ fits for the six stars.

Uncertainties in our LD diameter fits were derived using the method described in Tycner et al.\cite{2010SPIE.7734E.103T}, who showed that a non-linear least-squares method does not sufficiently account for atmospheric effects on time scales shorter than the window between target and calibrator observations. They describe a bootstrap Monte Carlo method that treats the observations as groups of data points because the NPOI collects data in scans consisting of 16 channels simultaneously. They discovered that when the 16 data points were analyzed individually, a single scan's deviation from the trend had a large impact on the resulting diameter and uncertainty calculation. On the other hand, when they preserved the inherent structure of the observational data by using the groups of 16 channels instead of individual data points, the uncertainty on the angular diameter was larger and more realistic. This method makes no assumptions about underlying uncertainties due to atmospheric effects, which are a problem for all stars observed using ground-based instruments. 

We combined our $\theta_{\rm LD}$ measurements with \emph{Hipparcos} parallaxes \cite{2007AandA...474..653V} to calculate the stars' $R$. In order to determine $L$ and $T_{\rm eff}$, we created SED fits to photometric values published in Ljunggren $\&$ Oja\cite{1965ArA.....3..439L}, McClure $\&$ Forrester\cite{1981PDAO...15..439M}, Olsen\cite{1993AandAS..102...89O}, Jasevicius et al.\cite{1990VilOB..85...50J}, Golay\cite{1972VA.....14...13G}, H{\"a}ggkvist $\&$ Oja\cite{1970AandAS....1..199H}, Kornilov et al.\cite{1991TrSht..63....1K}, Eggen\cite{1968tcpn.book.....E}, Johnson et al.\cite{1966CoLPL...4...99J}, Cutri et al.\cite{2003tmc..book.....C}, and Gezari et al.\cite{1993cio..book.....G} as well as spectrophotometry from Glushneva et al.\cite{1983TrSht..53...50G}, Glushneva et al.\cite{1998yCat.3207....0G}, and Kharitonov et al.\cite{1997yCat.3202....0K} obtained via the interface created by Mermilliod et al.\cite{1997AandAS..124..349M}. The assigned uncertainties for the 2MASS infrared measurements are as reported in Cutri et al.\cite{2003tmc..book.....C}, and an uncertainty of 0.05 mag was assigned to the optical measurements. We determined the best fit stellar spectral template to the photometry from the flux-calibrated stellar spectral atlas of Pickles\cite{1998PASP..110..863P} using the $\chi^2$ minimization technique. The resulting SEDs provided each star's bolometric flux ($F_{\rm BOL}$) and allowed for the calculation of extinction ($A_{\rm V}$) with the wavelength-dependent reddening relations of Cardelli et al.\cite{1989ApJ...345..245C}.

We used our $F_{\rm BOL}$ values combined with the stars' distances to estimate $L$ using $L = 4 \pi d^2 F_{\rm BOL}$. We also combined the $F_{\rm BOL}$ with $\theta_{\rm LD}$ to determine each star's effective temperature by inverting the relation,
\begin{equation}
F_{\rm BOL} = {1 \over 4} \theta_{\rm LD}^2 \sigma T_{\rm eff}^4,
\end{equation}
where $\sigma$ is the Stefan-Boltzmann constant and $\theta_{\rm LD}$ is in radians. Table 3 lists the results.

Giant star masses were estimated using the PARAM stellar model\footnote{http://stev.oapd.inaf.it/cgi-bin/param$\_$1.0} from Girardi et al.\cite{2000AandAS..141..371G} with a modified version of the method described in da Silva et al.\cite{2006AandA...458..609D}. The input parameters for each star were its interferometrically-measured $T_{\rm eff}$, its $V$ magnitude from Mermilliod\cite{2006yCat.2168....0M}, and its parallax\cite{2007AandA...474..653V} along with the corresponding error for each value. The model used these inputs to estimate each star's age, mass, radius, $(B-V)_0$, and log~$g$ using the isochrones and a Bayesian estimating method, calculating the probability density function separately for each property in question. da Silva et al. qualify mass estimates as ``more uncertain'' than other properties, so the resulting masses listed in Table 3  should be viewed as a rough estimates only.

\clearpage


\begin{center}
Table 2. Observing Log and Stellar Properties \\
\vspace{0.01in}
\begin{longtable}{cccccccc} \hline\hline
\label{example}
\vspace{0.01in}
   & Other & $\#$ of & $\#$ of    & $V$   & Parallax & $F_{\rm BOL}$                      \\
HD & Name  & Nights  & datapoints & (mag) & (mas)    & (10$^{-8}$ erg s$^{-1}$ cm$^{-2}$) \\
\hline \endfirsthead
\caption[]{\emph{continued}} \\
\hline
   & Other & $\#$ of & $\#$ of    & $V$   & Parallax & $F_{\rm BOL}$                      \\
HD & Name  & Nights  & datapoints & (mag) & (mas)    & (10$^{-8}$ erg s$^{-1}$ cm$^{-2}$) \\
\hline
\multicolumn{8}{r}{\emph{continued on the next page}}
\endfoot
\hline \endlastfoot
\hline
90537  & $\beta$ LMi    &  7 &  281 & 4.21 &  21.19$\pm$0.50 &  68$\pm$2 \\
113226 & $\epsilon$ Vir & 22 & 3954 & 2.83 &  29.76$\pm$0.14 & 233$\pm$4 \\
161797 & $\mu$ Her      & 23 &  328 & 3.42 & 120.33$\pm$0.16 & 102$\pm$1 \\
181276 & $\kappa$ Cyg   & 11 & 1420 & 3.79 &  26.27$\pm$0.10 & 100$\pm$2 \\
188512 & $\beta$ Aql    &  6 &  133 & 3.72 &  73.00$\pm$0.20 &  97$\pm$1 \\
219615 & $\gamma$ Psc   &  8 &  627 & 3.70 &  23.64$\pm$0.18 & 112$\pm$2 \\
\hline
\end{longtable}
\end{center}
\begin{center}
Note: HD 90537 is a spectroscopic binary, which may be why the parallax error is larger than other targets with similar parallaxes.
\end{center}

\section{SUMMARY}

\hskip15pt We observed six stars from the PTPS giant exoplanet host star candidate sample with the NPOI in order to measure their angular diameters. Those new values led to physical radii when combined with \emph{Hipparcos} parallaxes and led to bolometric fluxes when combined with photometric values from the literature and an SED fit. We then calculated the stars' luminosities and temperatures, and used the latter to provide a first estimate of the stars' masses. We will continue to observe the 25 stars available to the NPOI in order to improve their mass determinations. This will ultimately help us pin down which stellar evolution models are the most accurate by comparing the models' predictions with our measurements, and those models can be applied to stars too faint or small to be observed interferometrically. The updated masses will also improve our knowledge of the orbital properties, surface temperature, and minimum masses of the planets.


\begin{figure}[h]
\begin{center}
\includegraphics[width=1.0\textwidth]{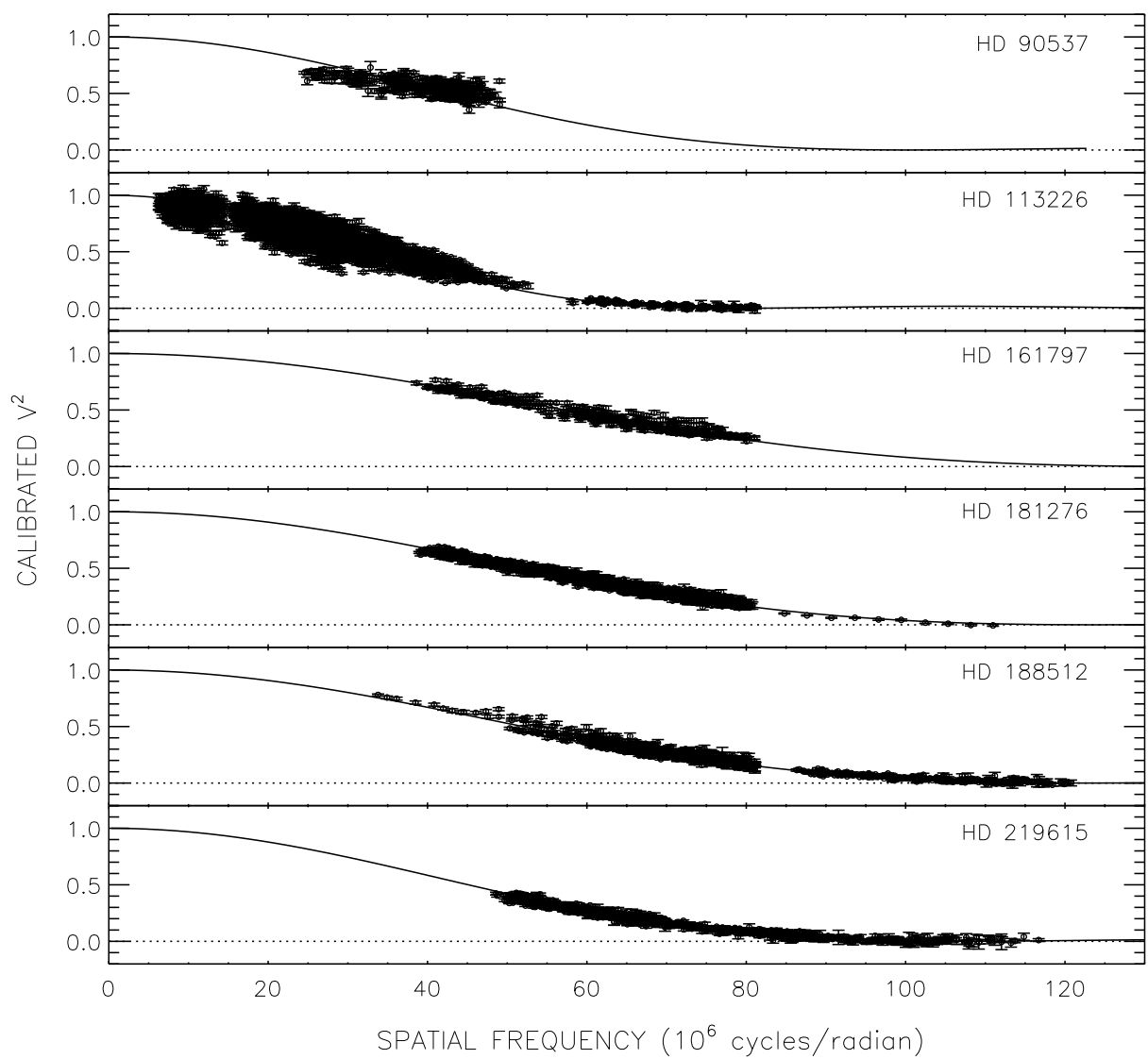}
\end{center}
\caption{$\theta_{\rm LD}$ fits for the six stars. The solid lines represent the visibility curve for the best fit $\theta_{\rm LD}$, the points are the calibrated visibilities, and the vertical lines are the measurement uncertainties.}
  \label{viscurves}
\end{figure}


\begin{center}
Table 3. Measured and Calculated Stellar Properties \\
\vspace{0.01in}
\begin{longtable}{cccccccc} \hline\hline
\label{example}
\vspace{0.01in}
   & $\theta_{\rm UD}$ &               & $\theta_{\rm LD}$ & $R$      & $M$         & $L$         & $T_{\rm eff}$ \\
HD & (mas)             & $\mu_\lambda$ & (mas)             & ($R_\odot$) & ($M_\odot$) & ($L_\odot$) & (K) \\
\hline \endfirsthead
\caption[]{\emph{continued}} \\
\hline
   & $\theta_{\rm UD}$ &               & $\theta_{\rm LD}$ & Radius      & $M$         & $L$         \\
HD & (mas)             & $\mu_\lambda$ & (mas)             & ($R_\odot$) & ($M_\odot$) & ($L_\odot$) \\
\hline
\multicolumn{8}{r}{\emph{continued on the next page}}
\endfoot
\hline \endlastfoot
\hline
90537  & 2.51$\pm$0.01 & 0.65 & 2.618$\pm$0.044 & 13.28$\pm$0.38 & 1.04$\pm$0.03 & 47.2$\pm$2.5 & 4149$\pm$41 \\
113226 & 3.18$\pm$0.01 & 0.65 & 3.321$\pm$0.023 & 11.99$\pm$0.10 & 2.57$\pm$0.01 & 82.3$\pm$1.5 & 5018$\pm$26 \\
161797 & 1.85$\pm$0.01 & 0.60 & 1.952$\pm$0.012 &  1.74$\pm$0.01 & 1.08$\pm$0.01 &  2.2$\pm$0.1 & 5324$\pm$17 \\
181276 & 2.03$\pm$0.01 & 0.69 & 2.172$\pm$0.005 &  8.89$\pm$0.04 & 2.10$\pm$0.08 & 45.3$\pm$0.8 & 5022$\pm$22 \\
188512 & 2.04$\pm$0.01 & 0.66 & 2.166$\pm$0.009 &  3.19$\pm$0.02 & 1.33$\pm$0.04 &  5.7$\pm$0.1 & 4992$\pm$11 \\
219615 & 2.35$\pm$0.01 & 0.65 & 2.481$\pm$0.011 & 11.28$\pm$0.10 & 2.11$\pm$0.16 & 62.7$\pm$1.5 & 4834$\pm$24 \\

\hline
\end{longtable}
\end{center}

\acknowledgments    
 
The Navy Precision Optical Interferometer is a joint project of the Naval Research Laboratory and the U.S. Naval Observatory, in cooperation with Lowell Observatory, and is funded by the Office of Naval Research and the Oceanographer of the Navy. This research has made use of the SIMBAD database, operated at CDS, Strasbourg, France. This publication makes use of data products from the Two Micron All Sky Survey, which is a joint project of the University of Massachusetts and the Infrared Processing and Analysis Center/California Institute of Technology, funded by the National Aeronautics and Space Administration and the National Science Foundation.


\bibliography{report} 
\bibliographystyle{spiebib} 

\end{document}